# Nonequilibrium Orbital Transitions via Applied Electrical Current in Calcium Ruthenate

Hengdi Zhao[1], Bing Hu[1,2], Feng Ye[3]*, Christina Hoffmann[3], Itamar Kimchi[1,4]* and Gang Cao[1]*

[1] Department of Physics, University of Colorado, Boulder, CO 80309, USA

[2] School of Mathematics and Physics, North China Electric Power University, Beijing 102206, China

[3] Neutron Scattering Division, Oak Ridge National Laboratory, Oak Ridge, TN 37831, USA

[4] JILA, NIST and Department of Physics, University of Colorado, Boulder, CO 80309, USA

Simultaneous control of structural and physical properties via applied electrical current poses a new, key research topic with both fundamental and technological significance. Studying the spin-orbit-coupled antiferromagnet $Ca_2RuO_4$, and its derivative with 3% Mn doping to alleviate the violent first-order transition at 357 K, we find that a small applied electrical current couples to the lattice by significantly reducing its orthorhombic distortions and octahedral rotations, while concurrently diminishing the 125 K- antiferromagnetic transition. Further increasing electrical current density above 0.15 A/cm$^2$ induces a new nonequilibrium orbital state, with a transition signature at 80 K that features a simultaneous jump in both magnetization and electrical resistivity, sharply contrasting the current-free state. We argue that nonequilibrium electron occupancies of the $t_{2g}$ orbitals stabilized by applied current drive the observed lattice changes and thereby the novel phenomena in this system. Finally, we note that current-induced diamagnetism reported in recent literature (*Science* **358**, 1084 (2017)) is not discerned in either slightly doped or pure $Ca_2RuO_4$.

*Corresponding authors: gang.cao@colorado.edu, yef1@ornl.gov, ikimchi@gmail.com

4d/5d-electron based oxides with inherent strong spin-orbit interactions (SOI) and significant electronic correlations create an entirely new hierarchy of energy scales and unique competitions between fundamental interactions. As a result, exotic quantum states arise whenever competing interactions conspire to generate large susceptibilities to small, external stimuli [1].

The antiferromagnetic (AFM) insulator $Ca_2RuO_4$ is a good example [2, 3]. With $Ru^{4+}(4d^4)$ ions, it exhibits a metal-insulator transition at $T_{MI}$ = 357 K [4], which marks a concomitant and particularly violent structural transition with a severe rotation and tilting of $RuO_6$. This structural transition removes the $t_{2g}$ orbital degeneracy ($d_{xy}, d_{yz}, d_{zx}$), dictating physical properties of $Ca_2RuO_4$ [4-15]. An AFM transition occurs only at a considerably lower Neel temperature $T_N$ =110 K [2,3], highlighting its close association with a further distorted structure. Extensive investigations of this system have established that quantum effects are intimately coupled to lattice perturbations [7, 12-19].

An early study demonstrates that electronic properties of $Ca_2RuO_4$ are sensitive to applied electrical current [21]. More recent investigations report current-induced diamagnetism [22] and current-induced non-equilibrium state [23]. It has become increasingly clear that electrical current as a new stimulus/probe controls quantum states in an unprecedented fashion. This is certainly manifested in our earlier study that demonstrates simultaneous current-control of structural and physical properties in the spin-orbit-coupled $Sr_2IrO_4$ [24].

In this work, we investigate structural, magnetic and transport properties as a function of electrical current and temperature in 3% Mn doped $Ca_2RuO_4$, and, for comparison, in pure $Ca_2RuO_4$ and 9% Mn doped $Ca_2RuO_4$. It is emphasized that the dilute Mn doping for Ru preserves the essential structural and physical properties of $Ca_2RuO_4$ but weakens the often pulverizing first-order structural phase transition at 357 K, making the single crystals more robust to sustain thermal cycling needed for thorough measurements [18, 19]. This work reveals that a novel coupling between *small* applied electrical current and the lattice drastically reduces the orthorhombic distortion and the octahedral rotation. The current-reduced lattice distortions in turn precipitously suppress the native AFM state and subsequently induce a nonequilibrium orbital state below 80 K that features a concurrent increase in both magnetization and electrical resistivity. The simultaneous measurements of both neutron diffraction and electrical resistivity provide a direct correlation between the current-reduced orthorhombicity and electrical resistivity. A temperature-current-density phase diagram generated based on the data illustrates a critical regime near a small current density, J, of 0.15 A/cm$^2$ that separates the native, diminishing AFM state and the emergent state. We argue that nonequilibrium electron occupancies of the $t_{2g}$ orbitals stabilized by applied current drive the critical lattice changes, thus the novel phenomena in this correlated, spin-orbit-coupled system. This study also emphasizes the conspicuous absence of current-induced diamagnetism, which is reported to exist in $Ca_2RuO_4$ [22]. Detailed experimental techniques are presented in the Supplemental Material (SM) [25]. We proceed with the discussion below focusing on current-induced changes in the magnetization, electrical resistivity and then the lattice modifications.

The magnetization along the *a* and *b* axis, $M_a$ and $M_b$, sensitively responds to applied current along the *b* axis. As demonstrated in **Figs.1a-1b**, the Néel temperature $T_N$ decreases systematically and rapidly from 125 K at current density J = 0 A/cm$^2$ to 29 K at J = 0.15 A/cm$^2$ in $M_b$ and 40 K at J = 0.12 A/cm$^2$ in $M_a$ and eventually vanishes at a critical current density $J_c$ ~ 0.15 A/cm$^2$ (the slight difference in $J_c$ for $M_a$ and $M_b$ is insignificant). The magnetic anisotropy between $M_a$ and $M_b$ with $M_a$ > $M_b$ is evident in **Figs.1a-1b**. Importantly, the diminishing AFM state is accompanied by a drastic decrease in the *b*-axis resistivity, $\rho_b$, by up to four orders of magnitude (**Fig.1c**), consistent with concurrent changes in orbital populations dictating the transport properties [7-15]. Note that $\rho_b$ and $M_b$ are simultaneously measured.

A new, distinct phase emerges as the AFM state vanishes. Let us focus on $M_a$ at higher J as $M_b$ behaves similarly. Immediately following the disappearance of the AFM, a pronounced anomaly marked by $T_O$ precipitates near $J_C$ (see **Fig.2**). The new transition temperature $T_O$ rises initially, peaks near J = 0.28 A/cm$^2$ before slowly decreasing with increasing J (**Fig.2a**). The simultaneously measured $\rho_b$ closely tracks $M_a$ with a well-defined anomaly corresponding to $T_O$, signaling a strong correlation between electron transport and magnetization in this emergent state (**Fig.2b**). The concurrent change in both $\rho_b$ and $M_a$ at $T_O$ sharply contrasts that of the native state in which $T_{MI} \gg T_N$ [4] and indicates a presence of a new type of state where magnetic response involves spin or orbital features or both.

The simultaneous rise in both $M_a$ and $\rho_b$ at $T_O$ would suggest a possible Slater transition to a new AFM state. However, this is inconsistent with other experimental observations. First, a sample history dependence above $T_O$ is seen in magnetization at higher current densities (J > 1 A/cm$^2$) [27]. Although the transition at $T_O$ is robust (no history dependence below $T_O$), a lack of history resetting above $T_O$ even after hours or days of equilibration time [27] implies that the transition at $T_O$ must involve changes not only in spins but also in orbital occupancies. Second, the disproportionally larger change in $\rho_b$ below $T_O$, compared to that in $M_a$, also indicates that spins alone cannot account for such a change in $\rho_b$ (**Fig.2a** and **2b**). For example, $\rho_b$ reduces by over three orders of magnitude below $T_O$ as J increases from 0.14 A/cm$^2$ to 2.8 A/cm$^2$ at 30 K (**Fig.2b inset**) whereas $M_b$ changes merely ~ 10% for the same J interval. All these contradicts a spin-driven Slater transition. Instead, it is consistent that the strong current produces a nonequilibrium state with inhomogeneous orbital occupancies. Such a current-induced state does not coexist with the native AFM state, implying an incompatible nature of the new state with the equilibrium, native state. Such an incompatibility is also true for other magnetic materials currently under investigation [27].

It is striking that current-induced diamagnetism, which is reported to exist in pure $Ca_2RuO_4$ [22], is absent in not only $Ca_2Ru_{0.97}Mn_{0.03}O_4$ but also pure $Ca_2RuO_4$. For comparison and clarification, we conduct the same measurements on pure $Ca_2RuO_4$ and 9% Mn doped $Ca_2RuO_4$. $M_a$ (and $\rho_b$, not shown) for pure $Ca_2RuO_4$ (**Fig. 2c**) and 9% Mn doped $Ca_2RuO_4$ (SM Fig.5 [25]) exhibits behavior remarkably similar to that seen in **Fig.2a** for 3% Mn doped $Ca_2RuO_4$. These results indicate the current-induced behavior above and below $T_O$ arises from the underlying properties of $Ca_2RuO_4$, independent of Mn doping. This is consistent with the fact that low Mn doping retains the underlying properties of $Ca_2RuO_4$ [19]. We also conduct a controlled study on the antiferromagnetic $BaIrO_3$ [26] whose magnetization has the same orders of magnitude; the results show no discernible current-induced changes in magnetization (SM Fig.3 [25]), ruling out any possible spurious behavior from our experimental setup.

We now turn to the current-driven crystal structure. $Ca_2Ru_{0.97}Mn_{0.03}O_4$ preserves the orthorhombic distortions (Pbca, No.61) of $Ca_2RuO_4$ [19]. The single crystal used for neutron diffraction with *in-plane* current applied is shown in **Fig.3a**. Two representative contour plots for the temperature dependence of the lattice parameters *a* and *b* at current density J = 0 and 4 A/cm$^2$ (**Fig.3b**) illustrate an abrupt change in the lattice parameters at $T_{MI}$ but no discernible *structural* inhomogeneity, which would lead to a broadening of Bragg peaks. (Note that this is different from nonequilibrium or inhomogeneous orbital occupancies discussed above.) Moreover, **Fig.3b** clearly shows a diminishing difference between the *a* and *b* axis or orthorhombicity with increasing J. Here we focus on the structural data at 100 K culled via neutron diffraction as a function of current density J applied within the basal plane. As shown in **Fig.3c**, one major effect is that the applied current progressively reduces the orthorhombicity, defined by $(b-a)/[(a+b)/2]$, from 4.4% at J = 0 A/cm$_2$ to 2.5% at J = 5 A/cm$_2$ and eventually to 1.2% at J = 30 A/cm$_2$ (**Fig.3d**). At the same time, the *c* axis expands by 1.2% and 2.4% at J = 5 A/cm$_2$ and 30 A/cm$_2$, respectively (**Fig.3e**). Interestingly, the current-driven lattice changes are remarkably similar to those due to modest pressure (< 2 GPa) [6]. The bond angle Ru-O-Ru, which defines the rotation of RuO$_6$ octahedra, increases by one degree at J = 5 A/cm$_2$ and two degrees at J = 18 A/cm$_2$, resulting in less buckled RuO$_6$ octahedra (**Figs.3f**). In addition, the bond angle O-Ru-O decreases from 91$_o$ to 90.2$_o$ at J = 5 A/cm$_2$, close to the ideal 90$_o$ (**Figs.3f**). Remarkably, the lattice changes more rapidly at smaller J (< 5 A/cm$_2$) than at larger J. In short, applied current significantly reduces the orthorhombic distortion from the tetragonal symmetry, expands the *c* axis and relaxes the bond angles, as schematically illustrated in **Figs.3g-3i**.

A direct link between the current-reduced orthorhombicity and electrical resistivity is further revealed in **Fig.4**. The contour plot of the orthorhombicity as functions of temperature and current density ranging from 0 to 4 A/cm$_2$ in **Fig. 4a** shows that the crystal structure sensitively responds to even a small current density J (< 1 A/cm$_2$). Particularly, the orthorhombicity is significantly reduced with increasing J throughout the measured temperature range. This change almost perfectly traces that of the electrical resistivity that was simultaneously measured during the neutron diffraction measurements, as illustrated in **Fig.4b**. This comparison provides an explicit correlation between the current-driven lattice and transport properties – The applied current reduces the orthorhombicity, modifying the $t_{2g}$ orbital occupancies that favor electron hopping. (Note that the structural transition at T$_{MI}$ (blue area in **Fig.4a**) barely shifts with J, confirming insignificant heating effect.)

Indeed, the more metallic state is a current-driven instability of an insulating state that is captured by local orbital occupancies. The experimental feature to explain is that applied in-plane current, as it suppresses $T_N$, also suppresses the transition to the insulator and the octahedral tilt and rotation while reducing the orthorhombicity and elongating the *c* axis (**Fig.3** and **Figs.4a-4c**). We now proceed to construct a theoretical framework that captures this feature through local orbital occupancies and overlaps. Without applied current, $Ru^{4+}$ ions nominally have 2 holes in the $t_{2g}$ orbitals but x-ray spectroscopy studies [1] suggest that a ½-hole is transferred to the oxygen. At high temperatures, in the metallic state, the remaining 3/2 hole is equally split in a 1:1 ratio between the $d_{xy}$ orbital and the manifold of $d_{xz}/d_{xz}$ orbitals (giving an electron occupancy roughly ~ $d_{xy}1.2d_{xz}1.6d_{yz}1.6$). At T < $T_{MI}$, the first-order structural transition at $T_{MI}$ = 357 K leads to the lattice distortions and the rotation, tilting and flattening of $RuO_6$, which transfers more holes from $d_{xy}$ to $d_{xz}/d_{yz}$, leading to a 1:2 ratio of hole occupancies in $d_{xy}$ vs $d_{xz}/d_{yz}$ (giving an electron occupancy roughly ~ $d_{xy}1.5d_{xz}1.5d_{yz}1.5$) [1]. The insulating state below $T_{MI}$ thus has each orbital at exactly 3/4 electron filling (or 1/4 hole filling). In contrast, the metallic state above $T_{MI}$ has unequal filling, with a nearly filled $d_{xz}/d_{yz}$ manifold (fewer holes) and, importantly, a nearly half-filled $d_{xy}$ orbital (more holes) (right sketch in **Fig.4d**). This analysis suggests that the conductivity in the metallic phase above $T_{MI}$ [11] is primarily enhanced by the $d_{xy}$ orbitals.

Now consider the nonequilibrium electron occupancies stabilized with an applied electric current. Within the $d_{xy}$ band, the electrons have large hopping amplitude from each Ru ion to each 4 of its neighbors, via the $p_x$ and $p_y$ orbitals on the four surrounding oxygens. This is not true for the $d_{xz}$ or $d_{yz}$ bands. So half-filling the $d_{xy}$ band is far more favorable for the conductivity than half filling either the $d_{xz}$ or $d_{yz}$ bands or uniformly quarter-filling the entire multi-band manifold.

Driving an in-plane current forces a metallic state to persist which, based on this picture, should lead to two effects: **(1)** the applied current minimizes crystalline distortions in the basal plane, so as to maximize inter-orbital hopping for in-plane conductivity; and **(2)** applying a current keeps the $d_{xy}$ band as close to half filling as possible, hence also avoiding the crystal distortions that are known (from the metal-insulator transition at zero applied current) to force $d_{xy}$ away from half filling. These two effects (**Figs.4b-4c**) capture the experimentally observed behavior of the resistivity and crystal structure with applied current while $T_N$ is suppressed. These current-induced lattice changes then also explain the vanishing native AFM state with increasing J because the AFM state requires a combination of rotation, tilt and flattening of $RuO_6$ octahedra hosting localized electrons [7, 10, 11], all of which are significantly weakened by applied current.

A temperature-current-density phase diagram generated based on the data presented above illustrates the diminishing AFM and the emergent, nonequilibrium order state characterized by $T_O$ with increasing J, as shown in **Fig.5**. The critical regime near $J_C$ = 0.15 A/$cm_2$ separates the two states and hosts a nonmagnetic state that is stabilized by delicate lattice changes discussed above. A spin-orbit-driven singlet $J_{eff}$ = 0 state [28] is suggested but more fine-tuning of the lattice in this regime is needed to rule in or out this possibility. For J > $J_C$, current-induced inhomogeneous orbital occupancies may play a more dominant role dictating physical properties both above and below $T_O$. All in all, at the heart of the new phenomena are the critical lattice modifications via current-driven nonequilibrium electron populations of the $t_{2g}$ orbitals.

**Acknowledgement:** This work is supported by NSF via grants DMR 1712101 and DMR 1903888. Work at ORNL was sponsored by the Scientific User Facilities Division, Basic Energy Sciences, U.S. Department of Energy (DOE). I.K. was supported by a National Research Council

Fellowship through the National Institute of Standards and Technology. We thank Dr. Daniel Khomskii for useful discussions.

**Captions**

**Fig.1. Current-driven magnetic and transport properties of $Ca_2Ru_{0.97}Mn_{0.03}O_4$:** The temperature dependence at various J applied along the *b* axis of **(a)** the *a*-axis magnetization $M_a$, **(b)** the *b*-axis magnetization $M_b$ and **(c)** the *b*-axis resistivity $\rho_b$. The magnetic field is at 1 T.

**Fig.2. Current-induced ordered state at J ≥ 0.14 A/cm²:** The temperature dependence at various J applied along the *b* axis of **(a)** $M_a$ at 1 Tesla and **(b)** $\rho_b$ for $Ca_2Ru_{0.97}Mn_{0.03}O_4$; **Inset**: $\rho_b$ at 30 K as a function of J. **(c)** $M_a$ at 1 Tesla for pure $Ca_2RuO_4$ at a few representative J for comparison.

**Fig.3. The neutron diffraction and current-driven lattice changes in $Ca_2Ru_{0.97}Mn_{0.03}O_4$:** **(a)** The sample holder for neutron diffraction with applied current. **Inset:** The single-crystal $Ca_2Ru_{0.93}Mn_{0.03}O_4$ with the electrical leads for applied current. **(b)** Two representative contour plots for the temperature dependence of the lattice parameters *a,* and *b* axis at current density J = 0 and 4 A/cm² applied in the basal plane. Note that the orthorhombicity is significantly reduced with increasing J. The current density J dependence at 100 K of **(c)** the *a* and *b* axis, **(d)** the orthorhombicity defined by $(b-a)/[(a+b)/2]$, **(e)** the *c* axis and **(f)** the bond angles Ru-O-Ru (red, left scale) and O-Ru-O (blue, right scale). The schematics illustrating the current-induced lattice changes: **(g)** the reduced orthorhombicity, **(h)** the elongated unit cell and **(i)** the increased bond angles; the displayed values for J = 0 and J = 5 A/cm², respectively.

**Fig.4. Direct correlation between the orthorhombicity and the electrical resistivity of $Ca_2Ru_{0.97}Mn_{0.03}O_4$**: The temperature-current-density contour plots for **(a)** the orthorhombicity and **(b)** electrical resistivity. The schematics illustrating at T > $T_O$ **(c)** the current-driven elongation of $RuO_6$ and **(d)** corresponding changes in $t_{2g}$ orbital populations.

**Fig.5. (a) The T-J phase diagram** illustrates that the applied current drives the system from the native AFM state (purple) through the critical regime near 0.15 A/cm$_2$ (gray) to the current-induced, nonequilibrium orbital state (blue).

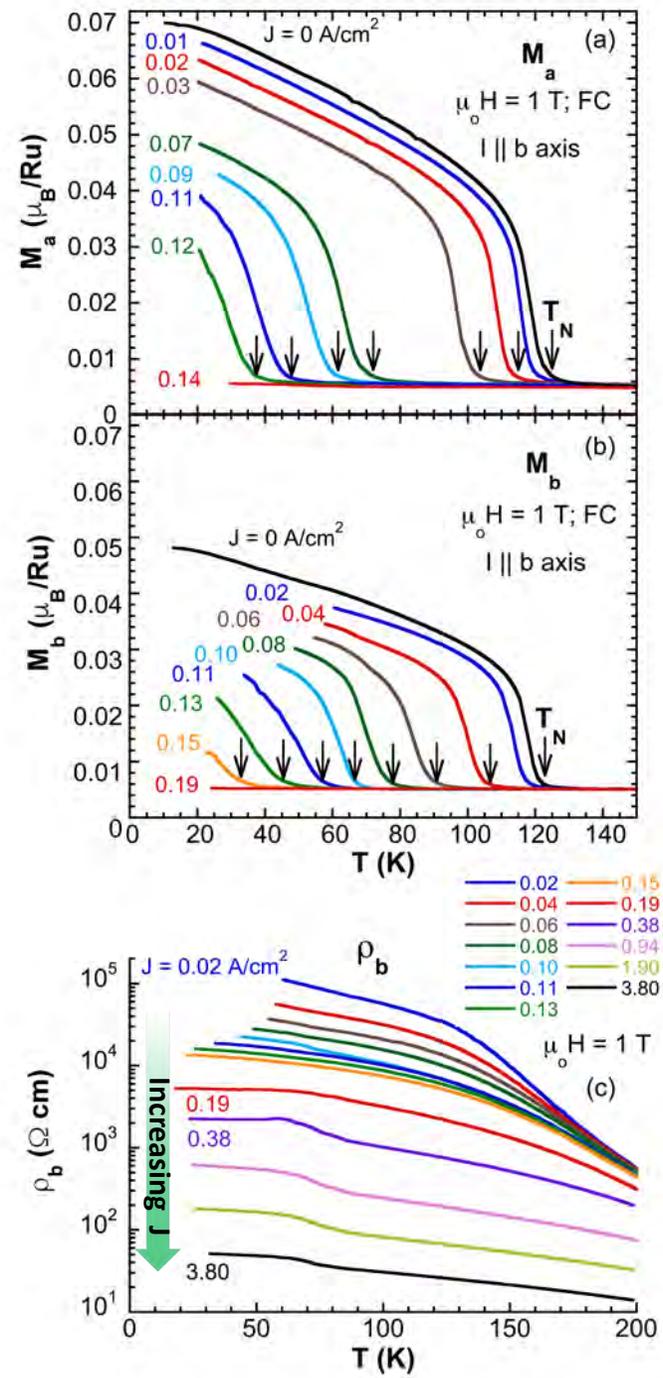

Fig.1

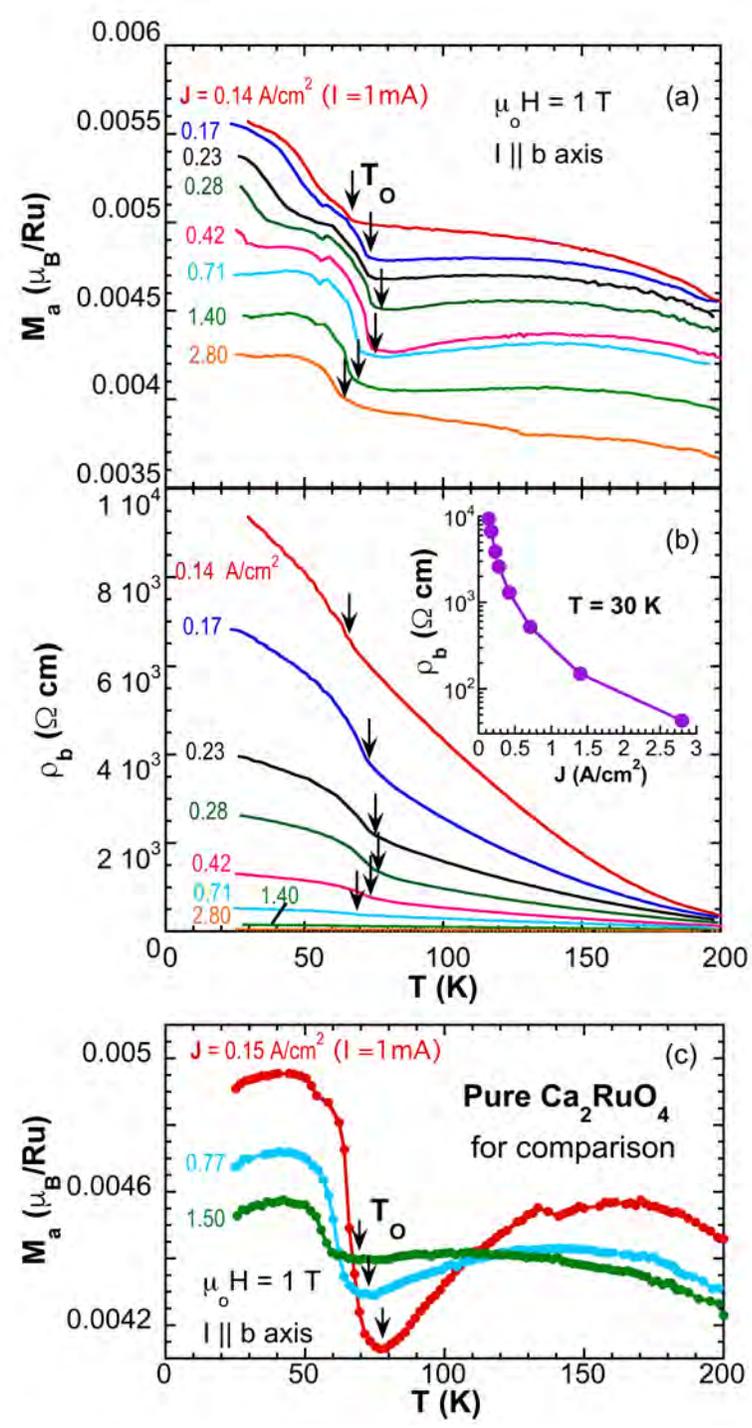

Fig.2

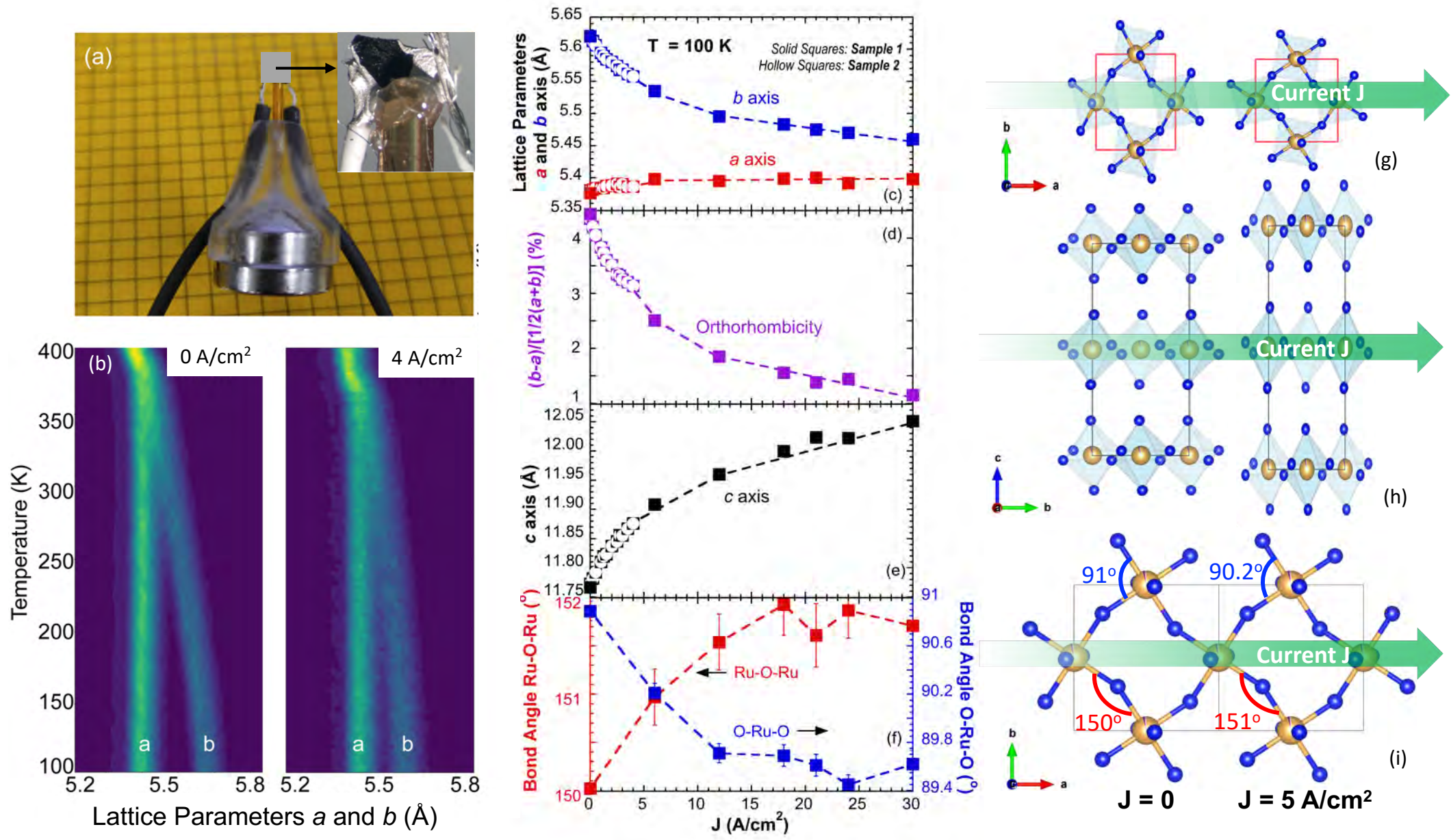

Fig.3

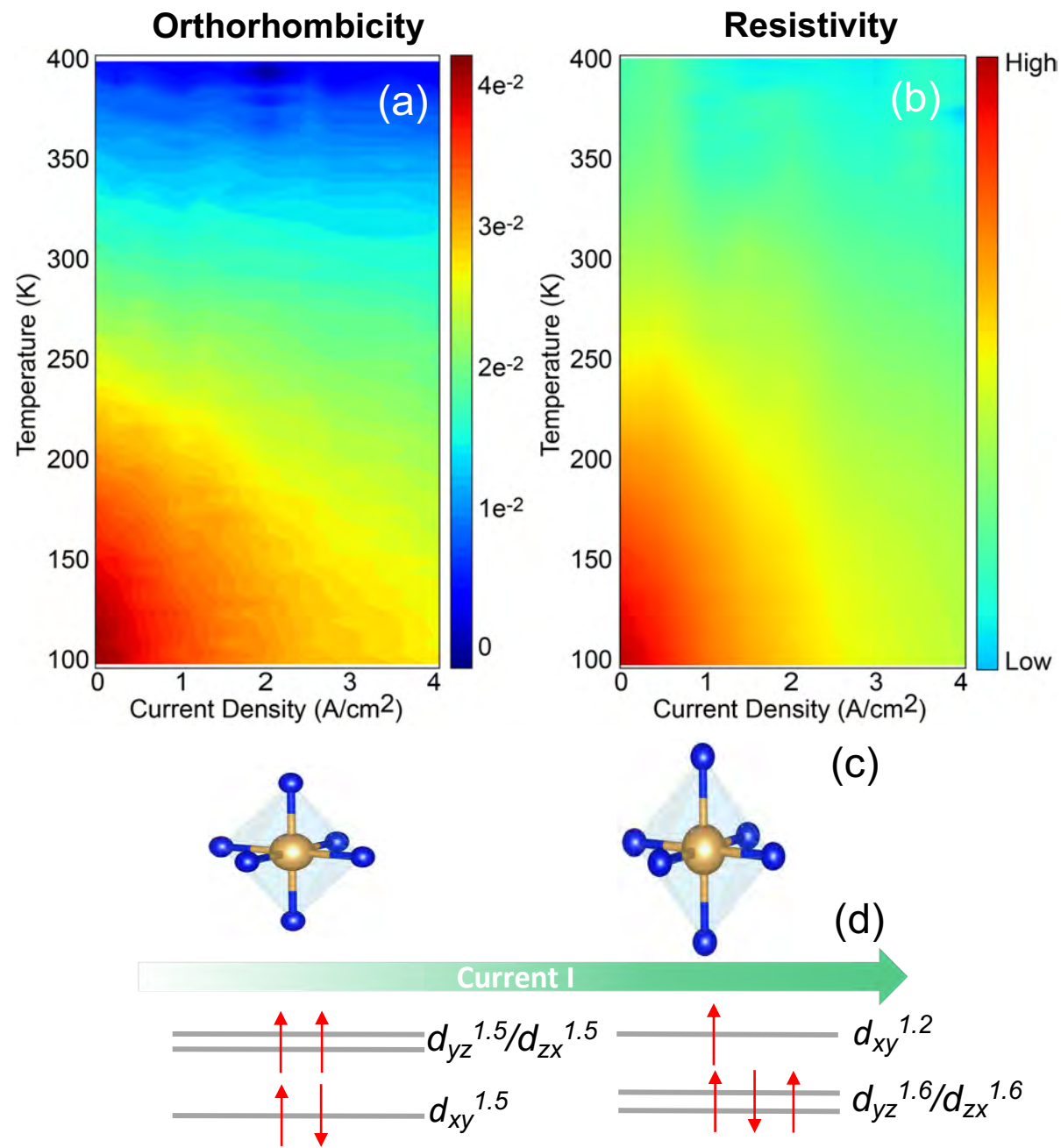

Fig.4

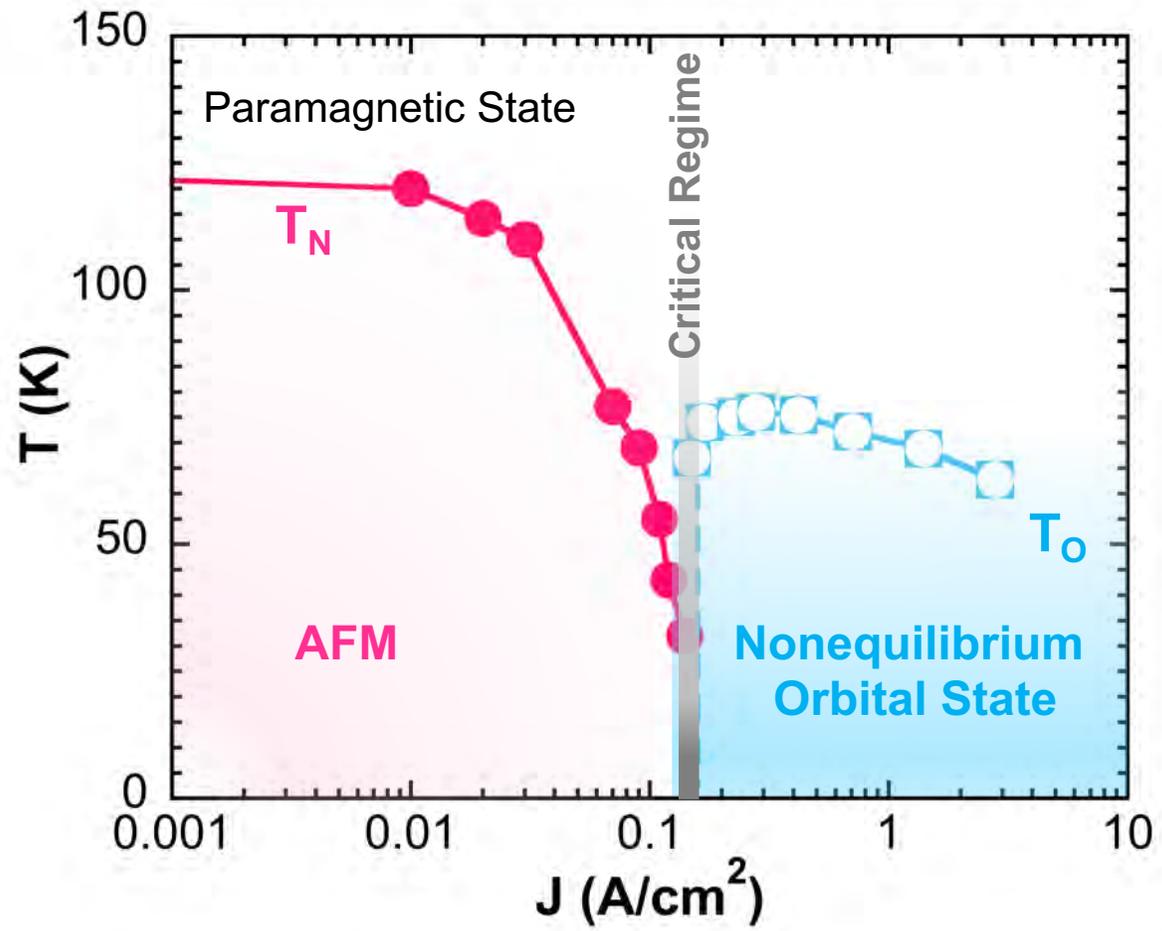

Fig.5